\documentclass[prl]{revtex4}%
\usepackage{amsfonts}
\usepackage{amsmath}
\usepackage{amssymb}
\usepackage{graphicx}%
\providecommand{\U}[1]{\protect\rule{.1in}{.1in}}

\begin{document}
\title{ Spectral Gap of the Anti-Ferromagnetic Lipkin-Meshkov-Glick Model}
\author{R. G. Unanyan}
\affiliation{Fachbereich Physik, Technische Universit\"{a}t Kaiserslautern, D-67663
Kaiserslautern, Germany}
\date{\today}

\begin{abstract}
The spectral property of the supersymmetric (SUSY) antiferromagnetic
Lipkin-Meshkov-Glick (LMG) model with an even number of spins is studied. The
supercharges of the model are explicitly constructed. By using the exact form
of the supersymmetric ground state we introduce simple trial variational
states for first excited states. It is demonstrated numerically that they
provide a relatively accurate upper bound for the spectral gap (the energy
difference between the ground state and first excited states) in all parameter
ranges. However, being an upper bound, it does not allow us to determine
vigorously whether the model is gapped or gapless. Here, we provide a
non-trivial lower bound for the spectral gap and thereby show that the
antiferromagnetic SUSY LMG model is gapped for any even number of spins.

\end{abstract}

\maketitle

\section{\bigskip Introduction}

About fifty years ago, an exactly solvable model of interacting fermions was
introduced in nuclear physics by Lipkin, Meshkov and Glick \cite{Lipkin}. This
model has been subsequently found widespread use not only in nuclear physics,
but also to a variety of other fields of physics, such as Bose--Einstein
condensates \cite{Cirac}, ion traps \cite{Unanyan2003} and cavities
\cite{Larson}. The Lipkin, Meshkov and Glick (LMG) Hamiltonian reads%
\begin{equation}
H=\xi\left(  \chi_{1}^{2}J_{x}^{2}+\chi_{2}^{2}J_{y}^{2}+\lambda\chi_{1}%
\chi_{2}J_{z}\right)  , \label{LMG_Hamiltonian}%
\end{equation}
where $J_{i}=\frac{1}{2}%
{\displaystyle\sum\limits_{n=1}^{N}}
\sigma_{i}^{\left(  n\right)  }$, $i=x,y,z$, are the familiar angular momentum
operators, $\sigma_{i}^{\left(  n\right)  }$ are the Pauli matrices, $N$ is
the total spin number. In the present paper, we focus our attention on the
manifold with maximum angular momentum $J=N/2$, where $N$ is an even integer.
So the Hilbert space has dimension $2J+1$. The parameters $\chi_{1,2},$ and
$\lambda$ are assumed to be positive constants. At the point where $\chi
_{2}=0$, the ground state of the Hamiltonian (\ref{LMG_Hamiltonian}) is either
ferromagnetic or antiferromagnetic depending on the sign of $\xi$. In the
present paper we discuss the antiferromagnetic case ($\xi>0$) and without loss
of generality, we set $\xi=1$. We notice that the rotation \ $\exp\left(
i\frac{\pi}{2}J_{x}\right)  $ transforms the Hamiltonian
(\ref{LMG_Hamiltonian}) into the form $\chi_{1}^{2}J_{y}^{2}+\chi_{2}^{2}%
J_{x}^{2}+\lambda\chi_{1}\chi_{2}J_{z}$ \ and therefore, we may assume that
$\chi_{1}\geq\chi_{2}$.\ 

As has been observed in refs.\cite{Unanyan2003} the spectrum of the model
(\ref{LMG_Hamiltonian}) at $\lambda=1$ presents a two-fold degeneracy in the
excited spectrum and non degenerate ground state with zero energy. These
observations manifest the presence of the supersymmetry (SUSY) in the system.
As far as the author knows, there is still no mathematical proof that the
Hamiltonian (\ref{LMG_Hamiltonian}) is supersymmetric and existence of a
non-vanishing spectral gap (the energy difference between the ground state and
first excited states) in the spectrum at $\lambda=1$. The aim of this paper is
to present a detailed proof of these observations.

First, we show that the LMG model at $\lambda=1$ is indeed supersymmetric by
constructing the supercharges $Q_{1}$ and $Q_{2}$
\begin{equation}
H=Q_{1}^{2}=Q_{2}^{2}, \label{charges}%
\end{equation}
and show that they anticommute, that is
\begin{equation}
\left\{  Q_{1},Q_{2}\right\}  =0. \label{SUSY_hamiltonian}%
\end{equation}
For a good introduction to the SUSY the reader is referred to review articles
e.g. \cite{Witten}. One consequence of this result is that it automatically
yields the two-fold degeneracy of excited states \cite{Witten}. Analogous
results are known as a result of elaborate approximative WKB computations
\cite{Garg}.

Second, we obtain upper and lower bounds for the spectral gap $\Delta E$ i.e.
the first excited state energy (since the ground state energy is zero) of the
Hamiltonian (\ref{LMG_Hamiltonian}). We show that a knowledge of the ground
state allows us to obtain a reasonable upper bound for the spectral gap by
using relatively simple variational states. The obtained bounds grow with the
system size. As a rule, the variational estimates are usually in good
agreement with exact eigenvalues of a Hamiltonian. Hence, it is natural to
expect that the true spectral gap of the antiferromagnetic SUSY LMG model
would also grow with $J$. We confirm this numerically and, moreover, provide a
non-trivial lower bound for the first excited state energy. So we rigorously
prove that the system (\ref{LMG_Hamiltonian}) at the supersymmetric point is
indeed gapped for any integer value of $J$.

\section{Supersymmetry in the LMG model}

\ The analysis of the spectrum of the Hamiltonian (\ref{LMG_Hamiltonian}) can
be greatly simplified by introducing a new set of variables $\Omega_{0}$ and
$\gamma$:
\begin{align}
\chi_{1}  &  =\Omega_{0}\cosh\gamma,\label{rotation}\\
\chi_{2}  &  =\Omega_{0}\sinh\gamma,\nonumber
\end{align}
where
\begin{equation}
\Omega_{0}^{2}=\chi_{1}^{2}-\chi_{2}^{2}\text{ \ and }\tanh\gamma=\frac
{\chi_{2}}{\chi_{1}}. \label{effectiv}%
\end{equation}
The Hamiltonian (\ref{LMG_Hamiltonian}) can be factorized by making use of the
identity
\begin{equation}
\exp\left(  -\gamma J_{z}\right)  J_{x}\exp\left(  \gamma J_{z}\right)
=J_{x}\cosh\gamma-iJ_{y}\sinh\gamma\label{rotationAngular}%
\end{equation}
for a hyperbolic rotation around the $z$ -axis with the parameter $\gamma$ as
follows%
\[
H=\Omega_{0}^{2}\left(  J_{x}\cosh\gamma+iJ_{y}\sinh\gamma\right)  \left(
J_{x}\cosh\gamma-iJ_{y}\sinh\gamma\right)  =
\]%
\begin{equation}
=\Omega_{0}^{2}\exp\left(  -\gamma J_{z}\right)  J_{x}\exp\left(  2\gamma
J_{z}\right)  J_{x}\exp\left(  -\gamma J_{z}\right)  .
\label{HamiltonianFacrorization}%
\end{equation}
It can easily be seen from Eq. (\ref{HamiltonianFacrorization}) that for
arbitrary values of $\gamma$ all eigenvalues of the Hamiltonian
(\ref{HamiltonianFacrorization}) are non-negative $E\geq0$.\ Notice that the
unitary transformation\ $\exp\left(  i\pi J_{x}\right)  $ changes the operator
$J_{z}\rightarrow-J_{z}$ and therefore the spectrum of the Hamiltonian \ is
symmetric with respect to $\gamma\rightarrow-\gamma$. Hence, it is sufficient
to restrict ourselves to the consideration of positive $\gamma$. One purpose
of this paper is to show that the LMG model (\ref{LMG_Hamiltonian}) at
$\lambda=1$ is supersymmetric. \ Instead of defining $Q_{1,2}$ operators in
terms of fermionic operators, we give their block-matrix representation in the
two invariant subspaces labeled by the "fermionic" number operator $F=\frac
{1}{2}\left(  {{1\!\!1}}+e^{i\pi\left(  J_{z}+J\right)  }\right)  $. \ The
operator $F$ has two distinct eigenvalues $0$ and $1$ with multiplicities
$J+1$ and $J$ respectively. It is easily seen that, the matrices
\begin{equation}
Q_{1}=\left(
\begin{array}
[c]{cc}%
\underset{\left(  J+1\right)  \times\left(  J+1\right)  }{\underbrace{0}} &
\underset{\left(  J+1\right)  \times\left(  J\right)  }{\underbrace{B}}\\
\underset{\left(  J\right)  \times\left(  J+1\right)  }{\underbrace{B^{\dagger
}}} & \underset{J\times J}{\underbrace{0}}%
\end{array}
\right)  ,\text{ \ \ \ }Q_{2}=\left(
\begin{array}
[c]{cc}%
\underset{\left(  J+1\right)  \times\left(  J+1\right)  }{\underbrace{0}} &
\underset{\left(  J+1\right)  \times\left(  J\right)  }{\underbrace{-iB}}\\
\underset{\left(  J\right)  \times\left(  J+1\right)
}{\underbrace{iB^{\dagger}}} & \underset{J\times J}{\underbrace{0}}%
\end{array}
\right)  , \label{SuperCharges}%
\end{equation}
where
\begin{equation}
B_{m_{1},m_{2}}=\Omega_{0}\left\langle m_{1}\right\vert \left(  J_{x}%
\cosh\gamma+iJ_{y}\sinh\gamma\right)  \left\vert m_{2}\right\rangle ,
\label{B_operator}%
\end{equation}%
\[
m_{1}=J,J-2,...-J,\text{ \ }m_{2}=J-1,J-3,...-\left(  J-1\right)
\]
fulfill the anticommutation relation (\ref{SUSY_hamiltonian}). The symbol
$\underset{\left(  M\right)  \times\left(  N\right)  }{\underbrace{X}}$
denotes a $M\times N$ matrix. We show that the relations (\ref{charges}) also
hold for these matrices. Indeed,
\[
Q_{1}^{2}=Q_{1}^{2}=\left(
\begin{array}
[c]{cc}%
\underset{\left(  J+1\right)  \times\left(  J+1\right)
}{\underbrace{BB^{\dagger}}} & \underset{\left(  J+1\right)  \times\left(
J\right)  }{\underbrace{0}}\\
\underset{\left(  J\right)  \times\left(  J+1\right)  }{\underbrace{0}} &
\underset{J\times J}{\underbrace{B^{\dagger}B}}%
\end{array}
\right)  ,
\]
and according to (\ref{B_operator}), for $m_{1},n_{1}=J,J-2,...-J$ we have
\[
\left(  BB^{\dagger}\right)  _{m_{1},n_{1}}=\Omega_{0}^{2}%
{\displaystyle\sum\limits_{m_{2}=J-1,...,-\left(  J-1\right)  }}
\left\langle m_{1}\right\vert \left(  J_{x}\cosh\gamma+iJ_{y}\sinh
\gamma\right)  \left\vert m_{2}\right\rangle \left\langle m_{2}\right\vert
\left(  J_{x}\cosh\gamma-iJ_{y}\sinh\gamma\right)  \left\vert n_{1}%
\right\rangle
\]
By inserting the resolution of identity%
\[%
{\displaystyle\sum\limits_{m=-\left(  J-1\right)  ,-\left(  J-3\right)
,...\left(  J-1\right)  }}
\left\vert m\right\rangle \left\langle m\right\vert ={1\!\!1-}%
{\displaystyle\sum\limits_{m=-J,-J+2,...J}}
\left\vert m\right\rangle \left\langle m\right\vert
\]
into the sum, we can write
\[
\left(  BB^{\dagger}\right)  _{m_{1},n_{1}}=\Omega_{0}^{2}\left\langle
m_{1}\right\vert \left(  J_{x}^{2}\cosh^{2}\gamma+J_{y}^{2}\sinh^{2}%
\gamma+J_{z}\sinh\gamma\cosh\gamma\right)  \left\vert n_{1}\right\rangle -
\]%
\[
-\Omega_{0}^{2}%
{\displaystyle\sum\limits_{m=J,J-2,...,-J}}
\left\langle m_{1}\right\vert \left(  J_{x}\cosh\gamma+iJ_{y}\sinh
\gamma\right)  \left\vert m\right\rangle \left\langle m\right\vert \left(
J_{x}\cosh\gamma-iJ_{y}\sinh\gamma\right)  \left\vert n_{1}\right\rangle =
\]%
\[
\Omega_{0}^{2}\left\langle m_{1}\right\vert \left(  J_{x}^{2}\cosh^{2}%
\gamma+J_{y}^{2}\sinh^{2}\gamma+J_{z}\sinh\gamma\cosh\gamma\right)  \left\vert
n_{1}\right\rangle \text{.}%
\]
In the last line we have used the fact $\left\langle m_{1}\right\vert \left(
J_{x}\cosh\gamma+iJ_{y}\sinh\gamma\right)  \left\vert m\right\rangle =0$, for
any $m_{1},m$ from the set $\left\{  J,J-2,...-J+2,-J\right\}  $.

\textit{\bigskip} Analogously, for any $m_{2},n_{2}$ from the set $\left\{
J-1,J-3,...-\left(  J-1\right)  \right\}  $ one has
\[
\left(  B^{\dagger}B\right)  _{m_{2},n_{2}}=\Omega_{0}^{2}\left\langle
m_{2}\right\vert \left(  J_{x}^{2}\cosh^{2}\gamma+J_{y}^{2}\sinh^{2}%
\gamma+J_{z}\sinh\gamma\cosh\gamma\right)  \left\vert n_{2}\right\rangle .
\]
Combining these two expressions the Hamiltonian (\ref{LMG_Hamiltonian}) in the
eigenbasis of $F$ can be partitioned into the block-diagonal form%

\[
H=\Omega_{0}^{2}\left(
\begin{array}
[c]{cc}%
\underset{\left(  J+1\right)  \times\left(  J+1\right)  }{\underbrace{J_{x}%
^{2}\cosh^{2}\gamma+J_{y}^{2}\sinh^{2}\gamma+J_{z}\sinh\gamma\cosh\gamma}} &
\underset{\left(  J+1\right)  \times\left(  J\right)  }{\underbrace{0}}\\
\underset{\left(  J\right)  \times\left(  J+1\right)  }{\underbrace{0}} &
\underset{J\times J}{\underbrace{J_{x}^{2}\cosh^{2}\gamma+J_{y}^{2}\sinh
^{2}\gamma+J_{z}\sinh\gamma\cosh\gamma}}%
\end{array}
\right)  .
\]
Hence, we have proved that the Hamiltonian (\ref{LMG_Hamiltonian}) is
supersymmetric at $\lambda=1$.

The factorized form (\ref{HamiltonianFacrorization}) allows us \ to write the
normalized ground state of (\ref{LMG_Hamiltonian}) in the following explicit
form
\begin{equation}
\left\vert \Psi_{g}\right\rangle =\frac{1}{\sqrt{P_{J}\left(  \cosh
2\gamma\right)  }}\exp\left(  \gamma J_{z}\right)  \left\vert m_{x}%
=0\right\rangle , \label{Ground_State}%
\end{equation}
with $P_{J}\left(  x\right)  $ being Legendre polynomials. Where, $\left\vert
m_{i}=m\right\rangle $ , $i=x,y,z$ denotes eigenvector of $J_{i}$ associated
with the magnetic quantum number $m$. For the excited states, however, no
further information can be gotten from the SUSY algebra (\ref{charges}) and
(\ref{SUSY_hamiltonian}), besides that they are two-fold degenerate. However,
as we will show below the explicit form of the ground state can be used to
find an accurate upper bound on the spectral gap.

\section{Upper bounds for the spectral gap}

When the ground state is known, an upper bound to the spectral gap $\Delta E$,
can be found from the following inequality \cite{Eckert}
\begin{equation}
\Delta E\leq\frac{\left\langle \Phi\right\vert H\left\vert \Phi\right\rangle
}{1-\left\vert \left\langle \Phi\right\vert \left.  \Phi_{g}\right\rangle
\right\vert ^{2}}, \label{Upper_bound}%
\end{equation}
where $\left\vert \Phi_{g}\right\rangle $ \ and $\left\vert \Phi\right\rangle
$ are normalized ground and trial wave functions respectively.

For purposes of calculation it is convenient to transform the Hamiltonian
(\ref{HamiltonianFacrorization}) into a unitarily equivalent
\begin{align}
H  &  =\Omega_{0}^{2}\left(  J_{z}^{2}\cosh^{2}\gamma+J_{y}^{2}\sinh^{2}%
\gamma-J_{x}\sinh\gamma\cosh\gamma\right)  =\label{Z_hamiltonian}\\
&  \Omega_{0}^{2}\exp\left(  -\gamma J_{x}\right)  J_{z}\exp\left(  2\gamma
J_{x}\right)  J_{z}\exp\left(  -\gamma J_{x}\right)  ,\nonumber
\end{align}
form.

Guided by our physical intuition, we choose the following trial state vector
\begin{equation}
\left\vert \Phi_{0}\right\rangle =\frac{\exp\left(  -\gamma J_{x}\right)
}{\sqrt{\left\langle +\right\vert \exp\left(  -2\gamma J_{x}\right)
\left\vert +\right\rangle }}\left\vert +\right\rangle , \label{Trial1}%
\end{equation}
where
\begin{equation}
\left\vert +\right\rangle =\frac{1}{\sqrt{2}}\left(  \left\vert m_{z}%
=1\right\rangle +\left\vert m_{z}=-1\right\rangle \right)  .
\label{Plus_Minus}%
\end{equation}

This choice is dictated by the following circumstances: First, for small and
large $\gamma$, it coincides with the first excited state of the Hamiltonian
(\ref{Z_hamiltonian}). Indeed, it is straightforward to show that the
Hamiltonian (\ref{Z_hamiltonian}) for these two limiting cases becomes simple
form%
\begin{equation}
H\approx\left\{
\begin{array}
[c]{c}%
\Omega_{0}^{2}J_{z}^{2}\text{ \ if }\gamma\rightarrow0\\
\Omega_{0}^{2}\frac{\exp\left(  2\gamma\right)  }{4}\left(  J(J+1-J_{x}\left(
J_{x}+1\right)  \right)  \text{ if }\gamma\rightarrow\infty.
\end{array}
\right.  \label{Simple_hamiltonian}%
\end{equation}
It is clear from (\ref{Trial1}) that for large $\left(  \delta=\gamma
J>>1\right)  $ the state $\left\vert \Phi_{0}\right\rangle $ coincides with
the first excited state $\left\vert m_{x}=-J\right\rangle $ of
(\ref{Simple_hamiltonian}). This is because, the operator $\exp\left(  -\gamma
J_{x}\right)  $ for large $\delta$ projects any state onto $\left\vert
m_{x}=-J\right\rangle $. While, for small $\delta$, the state $\left\vert
\Phi_{0}\right\rangle $ coincides with $\left\vert +\right\rangle $, which is
obviously a \ possible eigenstate for the first excited state of
(\ref{Simple_hamiltonian}).

Second, the vector (\ref{Trial1}) is orthogonal to the ground state vector
\begin{equation}
\left\vert \Phi_{g}\right\rangle =\frac{1}{\sqrt{P_{J}\left(  \cosh
2\gamma\right)  }}\exp\left(  +\gamma J_{x}\right)  \left\vert m_{z}%
=0\right\rangle . \label{Ground_State_z}%
\end{equation}
Although, this property of trial states is inessential for estimating the
spectral gap by the inequality (\ref{Upper_bound}). The orthogonality between
(\ref{Trial1}) and (\ref{Ground_State_z}) states allows us to simplify the
calculations of $\Delta E$.

It is worth noticing that the state
\begin{equation}
\left\vert \Phi_{V}\left(  \eta\right)  \right\rangle \sim\left(  \exp\left(
-\gamma J_{x}\right)  +\eta\right)  \left\vert +\right\rangle
\label{Trial_wave}%
\end{equation}
is also a suitable trial state vector for arbitrary values of $\eta$. We will
see below, by comparison with numerical simulations, that by a proper choice
of the parameter $\eta$ the states (\ref{Trial_wave}) and (\ref{Trial1})
produce almost the same upper bound for $\Delta E$.

\subsection{Fidelity}

\begin{figure}[ptb]%
\centering
\includegraphics[
height=1.9095in,
width=3.0459in
]%
{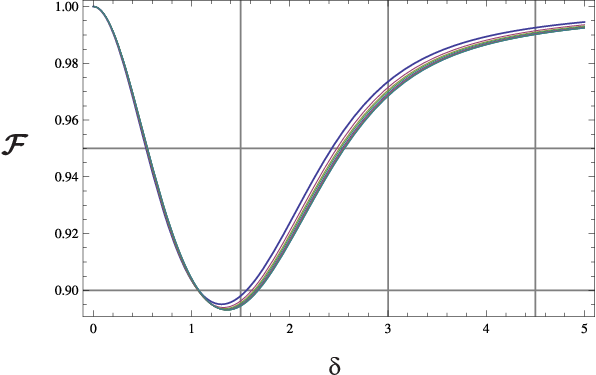}%
\caption{Fidelity as a function of $\delta=\gamma J$, for different values of
$J$, ($J=10,...100)$. For large $J$ the curves become virtually
indistinguishable.}%
\label{Fig2}%
\end{figure}

To quantify the exactness of the trial vector (\ref{Trial1}) we calculate the
fidelity $\mathcal{F}$ depending on the parameter $\delta=\gamma J$ for
different values of $J$. The fidelity, is simply the modulus of the overlap
between the state (\ref{Trial1}) and the exact (numerical) excited state of
the Hamiltonian (\ref{Z_hamiltonian})
\begin{equation}
\mathcal{F}=\left\vert \left\langle \Phi_{\mathtt{exact}}\right.  \left\vert
\Phi_{0}\right\rangle \right\vert . \label{Fidelity}%
\end{equation}
As we have seen, the first excited states are degenerate due to the SUSY. By
maximizing the fidelity over all possible superposition states with the same
energy, the fidelity $\mathcal{F}$ is shown in Fig.\ref{Fig2}, as a function
of $\delta=\gamma J$ for different values of $J$ ($J$ changes from $10$ to
$100$). Fig.\ref{Fig2} shows a generally high fidelity ($\mathcal{F\gtrsim
}0.89$) \ with a region of lower fidelity near $\delta=\gamma J\sim1$. The
exactness of (\ref{Trial1}) for the large and small $\delta$ is not surprising
because the choice of the state (\ref{Trial1}) was done to do this. However,
in our surprise, the trial state also describes an exact excited state with
high fidelity for intermediate values \ of $\delta$, where the perturbation
theories are not applicable.%

Now we return to the discussion of the spectral gap in our system.

\subsection{Variational upper bound for the gap}

Having discussed the exactness of our trial vector (\ref{Trial1}), we now
compare the expectation value of the energy $\Delta E_{0}=\left\langle
\Phi_{0}\right\vert H\left\vert \Phi_{0}\right\rangle $ of the state
(\ref{Trial1}) to the exact excited state energy by calculating the ratio
$\Delta E/\Delta E_{0}$. To express $\left\langle \Phi_{0}\right\vert
H\left\vert \Phi_{0}\right\rangle $ in suitable form we use the identity%
\begin{equation}
\left[  \exp\left(  -\gamma J_{y}\right)  \right]  _{m,m^{\prime}%
}=d_{m^{\prime},m}^{J}\left(  i\gamma\right)  , \label{Wigner}%
\end{equation}
where $d_{m^{\prime},m}^{J}\left(  x\right)  $ is the Wigner rotation matrix,
familiar from the quantum mechanics \cite{biedenharn}. After a lengthy but
straightforward calculation we eventually obtain the following expression
\begin{equation}
\Delta E_{0}=\Omega_{0}^{2}\left(  \cosh2\gamma+\left(  J+2\right)
\frac{P_{J-2}^{1,3}\left(  \cosh2\gamma\right)  \sinh\gamma\cosh^{3}%
\gamma+P_{J+2}^{-3,-1}\left(  \cosh2\gamma\right)  \sinh^{-3}\gamma\cosh
^{-1}\left(  \gamma\right)  }{P_{J-1}^{0,2}\left(  \cosh2\gamma\right)
\cosh^{2}\gamma+P_{J-1}^{2,0}\left(  \cosh2\gamma\right)  \sinh^{2}\gamma
}\sinh2\gamma\right)  , \label{first}%
\end{equation}
where $P_{n}^{\alpha,\beta}\left(  x\right)  $ are Jacobi polynomials
\cite{Szego}.

\begin{figure}[ptb]%
\centering
\includegraphics[
height=2.0652in,
width=3.2655in
]%
{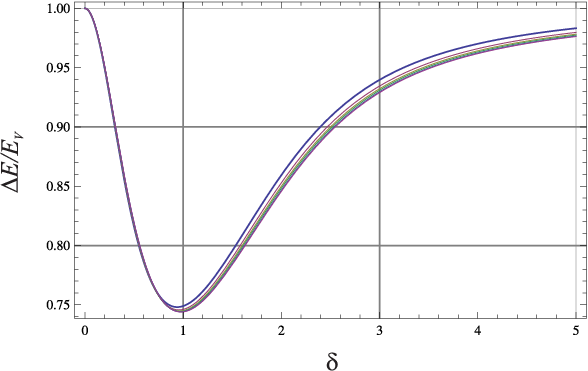}%
\caption{Energy of first excited states \ as a function of $\delta=\gamma J$
with respect to the variational bound (\ref{first}) for different $J$. }%
\label{Fig1}%
\end{figure}

To illustrate that the upper bound $\Delta E_{0}$ can be considered as an
accurate estimation of the gap, we have numerically calculated the first
excited state energy $\Delta E$ of the Hamiltonian. In Fig. \ref{Fig1}\ we
have plotted $\Delta E/\Delta E_{0}$ as a function of $\delta$ for for
different values of $J$ up to $100$.%

As we see from Fig.\ref{Fig1}\textbf{ }the behavior of $\Delta E/\Delta E_{0}$
is very similar to the fidelity $\mathcal{F}$. For sufficiently large values
of $J$ $\ $and $\gamma\neq0$, one can easily obtain a simple expression for
$\Delta E_{0}$. Making use of the asymptotics of the Jacobi polynomials for
large $J$, namely \cite{Szego},%
\begin{equation}
P_{J}^{\alpha,\beta}\left(  x\right)  \approx\frac{\left(  x^{2}-1\right)
^{-1/4}}{\sqrt{2\pi J}}\left[  x+\left(  x^{2}-1\right)  ^{1/2}\right]
^{J+1/2}\left(  x-1\right)  ^{-\alpha/2}\left(  x+1\right)  ^{-\beta/2}\left[
\left(  x+1\right)  ^{1/2}+\left(  x-1\right)  ^{1/2}\right]  ^{\alpha+\beta
},x>1 \label{Szego1}%
\end{equation}
one arrives at the simple expression%
\begin{equation}
\Delta E_{0}\approx\Omega_{0}^{2}\left(  \cosh2\gamma+J\sinh2\gamma\right)  .
\label{Upper_Limit}%
\end{equation}

\begin{figure}[ptb]%
\centering
\includegraphics[
height=1.8516in,
width=5.4604in
]%
{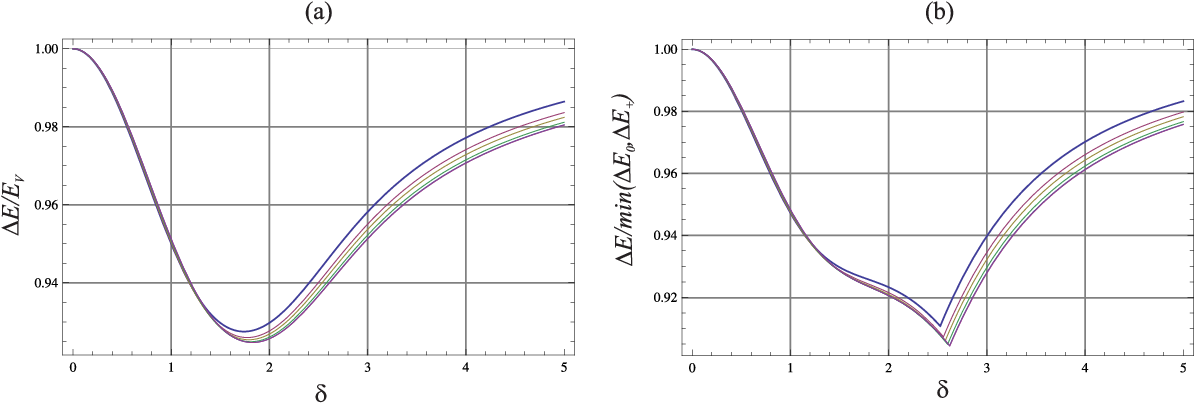}%
\caption{Energy of first excited states \ as a function of $\delta=\gamma J$
with resprct to the variational bounds obtained by minimization of $\Delta
E_{V}\left(  \eta\right)  $ (a) and $\min\left(  \Delta E_{0},\Delta
E_{+}\right)  $ (b) for different $J$ (from $10$ to $100$).}%
\label{Fig3}%
\end{figure}

Thus, we see $\Delta E_{0}$ grows linearly as $J>>1$ for all $\gamma>0$. A
better result for $\Delta E$ at intermediate values $\delta$ can be obtained
with the state $\left\vert +\right\rangle $. Substituting of the state vector
$\left\vert +\right\rangle $ in the formula (\ref{Upper_bound}), we obtain%
\begin{equation}
\Delta E_{+}=\frac{\left\langle +\right\vert H\left\vert +\right\rangle
}{1-\left\vert \left\langle +\right\vert \left.  \Phi_{g}\right\rangle
\right\vert ^{2}}=\Omega_{0}^{2}\frac{\cosh^{2}\gamma+\dfrac{\left(
J+2\right)  \left(  J-1\right)  }{4}\sinh^{2}\gamma}{1-\frac{2J}{J+1}%
\dfrac{P_{J}^{1,-1}\left(  \cosh\gamma\right)  }{P_{J}\left(  \cosh
2\gamma\right)  }} \label{Simple_plus}%
\end{equation}
Combining this expression with formula (\ref{first}), we can derive an upper
bound for
\begin{equation}
\Delta E\leq\Omega_{0}^{2}\min\left[  \Delta E_{0},\Delta E_{+}\right]  .
\label{betterBound0}%
\end{equation}
For a state composed of $\left\vert \Phi_{0}\right\rangle $ and $\left\vert
+\right\rangle $ states i.e. the state (\ref{Trial_wave}) the minimum of
\[
\Delta E_{V}\left(  \eta\right)  =\frac{\left\langle \Phi_{V}\left(
\eta\right)  \right\vert H\left\vert \Phi_{V}\left(  \eta\right)
\right\rangle }{\left\langle \Phi_{V}\left(  \eta\right)  \right\vert \left.
\Phi_{V}\left(  \eta\right)  \right\rangle -\left\vert \left\langle \Phi
_{V}\left(  \eta\right)  \right\vert \left.  \Phi_{g}\right\rangle \right\vert
^{2}}%
\]
would be better than (\ref{betterBound0}). We do not explicitly give this
lengthy optimal bound, but rather present in Fig.\ref{Fig3} the minimum of
$\Delta E_{V}\left(  \eta\right)  /\Delta E$ (see Fig.\ref{Fig3}a) and
$\min\left[  \Delta E_{0},\Delta E_{+}\right]  /\Delta E$ as function of
$\delta$.
As we can see from Fig.\ref{Fig3}(a), the variational state (\ref{Trial_wave})
which depends on the single parameter $\eta$, accurately recovers the true
excited state energy within $0.8\%$. \ The accuracy of the relatively simple
upper bound (\ref{betterBound0}) ( see Fig.\ref{Fig3}(b)) is a few percent
less than the optimal bound given by (\ref{Trial_wave}). It is a remarkable
result considering the simplicity of the trial function used. Thus, we have
seen that at a quantitative level the variational upper bounds for the gap are
in a good agreement with exact numerical results (for $J$ up $100$). But these
observations do not provide a rigorous proof that the system is indeed gapped.
So the problem is to get a non trivial lower bound for $\Delta E$. Next we
turn our attention to this issue.

\section{Lower bound for the spectral gap}

We now turn to the problem of obtaining a non trivial lower bound for the gap
by a suitable choice of the form of the Hamiltonian. To this end, we need some
facts from the theory of matrices. For notation simplicity, throughout this
section we denote by $\lambda_{n}\left(  X\right)  $ $\left(
n=1,2,...M\right)  $ eigenvalues of a $M\times M$ Hermitian matrix\ $X$ in
increasing order, i.e. $\left\{  \lambda_{1}\left(  X\right)  \geq\lambda
_{2}\left(  X\right)  \geq...\geq\lambda_{M}\left(  X\right)  \right\}  $. \ A
lower bound for the spectral gap can be obtained using the Weyl's theorem
\cite{Weyl} which is stated as follows. Let $X$, $Y$ \ and $Z$ with $Z=X+Y$ be
$M$-dimensional Hermitian matrices, then
\begin{equation}
\lambda_{i+j-1}\left(  Z\right)  \leq\lambda_{i}\left(  X\right)  +\lambda
_{j}\left(  Y\right)  , \label{Weyl11}%
\end{equation}
whenever $i+j-1\leq M$, or equivalently%
\begin{equation}
\lambda_{k}\left(  Z\right)  \geq\lambda_{i}\left(  X\right)  +\lambda
_{j}\left(  Y\right)  , \label{lower_inequality}%
\end{equation}
if $i+j=M+k$.

To apply the Weyl's theorem we split the Hamiltonian (\ref{Z_hamiltonian})
into two parts:
\begin{equation}
H=H_{X}+\Omega_{0}^{2}J_{z}^{2}, \label{Splitted_Hamiltonian}%
\end{equation}
where
\[
H_{X}=\Omega_{0}^{2}\left(  \left(  J_{z}^{2}+J_{y}^{2}\right)  \sinh
^{2}\gamma-J_{x}\sinh\gamma\cosh\gamma\right)  =
\]%
\begin{equation}
=\Omega_{0}^{2}\left(  \left(  J(J+1)-J_{x}^{2}\right)  \sinh^{2}\gamma
-J_{x}\sinh\gamma\cosh\gamma\right)  . \label{Hamiltonian_x}%
\end{equation}
In the second line we have used the identity $J_{z}^{2}+J_{y}^{2}+J_{x}%
^{2}=J(J+1)$. Using the inequality (\ref{lower_inequality}) and recalling
supersymmetric property of our Hamiltonian for the spectral gap, i.e. $\Delta
E=\lambda_{2J}\left(  H\right)  =\lambda_{2J-1}\left(  H\right)  $, one
obtains
\begin{equation}
\Delta E/\Omega_{0}^{2}\geq\lambda_{i}\left[  \left(  J(J+1)-J_{x}^{2}\right)
\sinh^{2}\gamma-J_{x}\sinh\gamma\cosh\gamma\right]  +\lambda_{j}\left(
J_{z}^{2}\right)  , \label{Inequalities}%
\end{equation}
where $i+j=4J$. It is not hard to verify directly that the possible pairs of
$\left(  i,j\right)  $ are $\left(  2J+1,2J-1\right)  ,$ $\left(
2J-1,2J+1\right)  ,$ and $(2J,2J)$. The corresponding energies are
\begin{equation}
E_{ge}/\Omega_{0}^{2}=1-J\exp\left(  \gamma\right)  \cdot\sinh\gamma
\rightarrow\left(  2J+1,2J-1\right)  , \label{Energies1}%
\end{equation}%
\begin{equation}
E_{eg}/\Omega_{0}^{2}=\min\left[  J\cdot\mathtt{\exp}\left(  \gamma\right)
\cdot\sinh\gamma,\left(  5J-4\right)  \sinh^{2}\gamma-\frac{1}{2}\left(
J-2\right)  \sinh2\gamma\right]  \rightarrow\left(  2J-1,2J+1\right)  ,
\label{Energy2}%
\end{equation}%
\begin{equation}
E_{ee}/\Omega_{0}^{2}=1+\left(  3J-1\right)  \sinh^{2}\gamma-\frac{1}%
{2}\left(  J-1\right)  \sinh2\gamma\rightarrow\left(  2J,2J\right)
\label{Energy3}%
\end{equation}
The bound (\ref{Energies1}) is a trivial one, it goes to negative values as
$\gamma$ becomes large $\gamma>\ln\left(  1+2/J\right)  \approx2/J$. For small
$\gamma<<1/J$ , the bound (\ref{Energy3}) is better than (\ref{Energy2}) and
it becomes negative for $1/J\lesssim\gamma\lesssim0.3$. While, for relatively
large $\gamma\gtrsim0.2$ the bound (\ref{Energy2}) is much better than
(\ref{Energy3}), although it becomes negative when $\gamma\lesssim0.2$. Hence,
on the basis of these analyses we arrive at the inequality for
\begin{equation}
\Delta E\geq\Omega_{0}^{2}\min\left[  J\cdot\mathtt{\exp}\left(
\gamma\right)  \cdot\sinh\gamma,\text{ }\left(  5J-4\right)  \sinh^{2}%
\gamma-\frac{1}{2}\left(  J-2\right)  \sinh2\gamma\right]  ,
\label{First_Estimation_Weyl}%
\end{equation}
which gives a trivial bound for $\gamma\lesssim0.2$ but a nontrivial one for
large $\gamma$. By comparing (\ref{Upper_Limit}) and
(\ref{First_Estimation_Weyl}) for large $\gamma$ and $J$ we conclude that the
upper and lower bounds on $\Delta E$ \ converge to each other. Hence, it
remains \ to be found a non trivial lower bound for intermediate values of
$\gamma$.

In the following we show that by using the representation
(\ref{HamiltonianFacrorization}) the degeneracy and a non trivial lower bound
(for arbitrary $\gamma$ and integer $J$), for $\Delta E$ can be obtained
through direct calculations.

It is easily seen that the Hamiltonian (\ref{HamiltonianFacrorization}) is
similar to the non-Hermitian Hamiltonian
\[
H_{\mathtt{n}}=\exp\left(  -\gamma J_{x}\right)  H\exp\left(  \gamma
J_{x}\right)  =\Omega_{0}^{2}\exp\left(  -2\gamma J_{x}\right)  J_{z}%
\exp\left(  2\gamma J_{x}\right)  J_{z}=
\]%
\begin{equation}
=\Omega_{0}^{2}\left(  J_{z}^{2}\cosh2\gamma+iJ_{y}J_{z}\sinh2\gamma\right)  .
\label{Non_Hermitian}%
\end{equation}
We show that, beside its non-Hermiticity the spectral properties of
(\ref{Non_Hermitian}) are very transparent. Indeed, from Eq.
(\ref{Non_Hermitian}), one can see clearly first that it has a null state
$\left\vert m_{z}=0\right\rangle $ ( $J_{z}\left\vert m_{z}=0\right\rangle =0$
). Second, all excited states are two-fold degenerate. To see this, we notice
that $H_{\mathtt{n}}$ can be represented as a block matrix
\begin{equation}
H_{\mathtt{n}}=\left(
\begin{array}
[c]{ccc}%
\underset{J\times J}{\underbrace{H_{-}}} & \underset{1\times J}{\underbrace{0}%
} & \underset{J\times J}{\underbrace{0}}\\
\left\langle a\right\vert  & 0 & \left\langle b\right\vert \\
\underset{J\times J}{\underbrace{0}} & \underset{1\times J}{\underbrace{0}} &
\underset{J\times J}{\underbrace{H_{+}}}%
\end{array}
\right)  , \label{BlockForm22}%
\end{equation}
in the eigenbasis of $J_{z}$. $H_{+}$ and $H_{-}$ are real permutation
equivalent matrices, i.e. they have the same spectrum. The transposed vectors
$\left\langle a\right\vert $ and $\left\langle b\right\vert $ connect the
state $\left\vert m_{z}=0\right\rangle $ with negative and positive magnetic
quantum numbers $m_{z}$.

As an example, consider $J=2$, the Hamiltonian $H_{\mathtt{n}}$ , has the
following form%

\[
H_{\mathtt{n}}=\Omega_{0}^{2}\left(
\begin{array}
[c]{ccccc}%
4\cosh2\gamma & \sinh2\gamma & 0 & 0 & 0\\
-2\sinh2\gamma & \cosh2\gamma & 0 & 0 & 0\\
0 & -\frac{\sqrt{6}}{2}\sinh2\gamma & 0 & -\frac{\sqrt{6}}{2}\sinh2\gamma &
0\\
0 & 0 & 0 & \cosh2\gamma & -2\sinh2\gamma\\
0 & 0 & 0 & \sinh2\gamma & 4\cosh2\gamma
\end{array}
\right)  ,
\]
where%
\[
H_{-}=\Omega_{0}^{2}\left(
\begin{array}
[c]{cc}%
4\cosh2\gamma & \sinh2\gamma\\
-2\sinh2\gamma & \cosh2\gamma
\end{array}
\right)  ,\text{ \ }H_{+}=\Omega_{0}^{2}\left(
\begin{array}
[c]{cc}%
\cosh2\gamma & -2\sinh2\gamma\\
\sinh2\gamma & 4\cosh2\gamma
\end{array}
\right)
\]
and%
\[
\left\langle a\right\vert =\Omega_{0}^{2}\left(  0,-\frac{\sqrt{6}}{2}%
\sinh2\gamma\right)  ,\text{ }\left\langle b\right\vert =\Omega_{0}^{2}\left(
-\frac{\sqrt{6}}{2}\sinh2\gamma,0\right)  .
\]
One can see that $H_{+}$ and $H_{-}$ are permutation equivalent, i.e.
\[
H_{-}=\left(
\begin{array}
[c]{cc}%
0 & 1\\
1 & 0
\end{array}
\right)  H_{+}\left(
\begin{array}
[c]{cc}%
0 & 1\\
1 & 0
\end{array}
\right)  .
\]

The matrix elements of $H_{+}$ \ are
\begin{align}
\left(  H_{+}\right)  _{m,m^{\prime}}  &  =\Omega_{0}^{2}\left(  m^{2}%
\delta_{mm^{\prime}}\cosh\left(  2\gamma\right)  +\frac{m^{\prime}}{2}%
\sinh2\gamma\left[  \delta_{m,m^{\prime}+1}\sqrt{\left(  J-m^{\prime}\right)
\left(  J+m^{\prime}+1\right)  }-\delta_{m,m^{\prime}-1}\sqrt{\left(
J+m^{\prime}\right)  \left(  J-m^{\prime}+1\right)  }\right]  \right)
\label{matrixelements_negative}\\
m,m^{\prime}  &  =J,J-1,...,1.\nonumber
\end{align}
The spectrum of $H_{\mathtt{n}}$ can be obtained from the following algebraic
equation
\begin{equation}
\lambda\det\left(  \lambda\cdot{{1\!\!1}}_{J\times J}-H_{+}\right)
\det\left(  \lambda\cdot{{1\!\!1}}_{J\times J}-H_{-}\right)  =0.
\label{determinants}%
\end{equation}
Since $H_{+}$ and $H_{-}$ have the same spectrum, the spectrum of
$H_{\text{n}}$ is doubly degenerate except for the eigenstate $\left\vert
m_{z}=0\right\rangle $. \ Therefore, we may restrict ourselves to the study of
spectral properties of $H_{+}$. We thus have verified directly that for
integer $J$\ and for arbitrary $\gamma$ the spectrum of the initial
Hamiltonian (\ref{HamiltonianFacrorization}) is supersymmetric. And, in
addition to that the excited spectrum of the Hamiltonian
(\ref{HamiltonianFacrorization}) coincides with the spectrum $H_{+}$ i.e. the
spectral gap of our model coincides with the ground state energy of $H_{+}$.

As one can see from Eq (\ref{matrixelements_negative}), the matrix $H_{+}$ can
be represented in a compact form
\begin{equation}
H_{+}=\Omega_{0}^{2}\left(  j_{z}^{2}\cosh2\gamma+i\cdot j_{y}\cdot j_{z}%
\sinh2\gamma\right)  , \label{short_angular_momentum}%
\end{equation}
where the truncated Hermitian angular momentum matrices $j_{z}$ , $j_{y}$ and
$j_{x}$ satisfy the following commutation relations%
\begin{align}
\left[  j_{z},j_{x}\right]   &  =ij_{y},\label{truncated_angular_momentum}\\
\left[  j_{y},j_{z}\right]   &  =ij_{x},\nonumber
\end{align}
but unlike the ordinary angular momentum operators
\begin{equation}
\left[  j_{x},j_{y}\right]  \neq ij_{z}. \label{x_y_z}%
\end{equation}
Next, we will show that the Hamiltonian $H_{+}$ can be transformed into a more
pleasant form. To this end, we recall that $j_{z}$ is a positive definite
matrix with eigenvalues $1,2,...J$. So that the square root $j_{z}^{1/2}$is
well defined. Using this, the matrix $H_{+}$ can be written as
\begin{equation}
H_{+}=j_{z}^{-1/2}\cdot h\cdot j_{z}^{1/2}, \label{second_transformation}%
\end{equation}
where
\begin{equation}
h=\Omega_{0}^{2}\left(  j_{z}^{2}\cosh2\gamma+i\cdot j_{z}^{1/2}\cdot
j_{y}\cdot j_{z}^{1/2}\sinh2\gamma\right)  \label{Hermition}%
\end{equation}
which has the same spectrum as $H_{+}$. Now we state that for any $\gamma$ and
$J$
\begin{equation}
\Delta E\geq\Omega_{0}^{2}\cosh2\gamma. \label{main_estimation}%
\end{equation}

Indeed, for any eigenvalue $E\left(  h\right)  >0$ of $h$ and its
corresponding normalized eigenvector $\left\vert \varphi\right\rangle $ we
have
\[
E\left(  h\right)  =\left\langle \varphi\right\vert h\left\vert \varphi
\right\rangle =
\]%
\[
=\Omega_{0}^{2}\left(  \left\langle \varphi\right\vert j_{z}^{2}\left\vert
\varphi\right\rangle \cosh2\gamma+i\left\langle \varphi\right\vert j_{z}%
^{1/2}\cdot j_{y}\cdot j_{z}^{1/2}\left\vert \varphi\right\rangle \sinh
2\gamma\right)  =
\]%
\[
=\Omega_{0}^{2}\left\langle \varphi\right\vert j_{z}^{2}\left\vert
\varphi\right\rangle \cosh2\gamma\geq
\]%
\[
\geq\Omega_{0}^{2}\cosh2\gamma
\]
where in the second line we have used the fact that $j_{z}^{1/2}\cdot
j_{y}\cdot j_{z}^{1/2}$ is a Hermitian matrix. In the last line we have used
the fact that the smallest eigenvalue of $j_{z}$ is equal $1$. Hence, all
eigenvalues of $h$ lie at finite distances from the origin greater than
$\cosh2\gamma$ i.e. the spectral gap of the antiferromagnetic SUSY LMG
Hamiltonian (\ref{HamiltonianFacrorization}) is bounded from below by
$\Omega_{0}^{2}\cosh2\gamma$. Unlike the inequality
(\ref{First_Estimation_Weyl}), it does not involve $J$.

\section{ Discussion and Conclusion}

In the present paper we have investigated the spectrum of the
antiferromagnetic LMG model at the SUSY point. We have proved, by explicitly
constructing the supercharges, that the Hamiltonian (\ref{LMG_Hamiltonian}) is
supersymmetric at $\lambda=1$. By using the explicit form \ of the ground
state of the Hamiltonian we have introduced variational excited states that
have pretty high fidelity with the exact excited state. A simple form of these
states enables us to find closed expressions for upper bounds for the spectral
gap. It was shown numerically that the obtained upper bounds are in good
agreement with the exact spectral gap. Simple but non trivial lower bounds for
the spectral gap was found. Thus, we have shown that the antiferromagnetic
SUSY LMG model is gapped for any values of $\gamma$. Although, for
intermediate values of $\gamma$ the obtained lower bound
\[
\Delta E\geq\Omega_{0}^{2}\max\left[  \cosh2\gamma,\min\left[  J\cdot
\mathtt{\exp}\left(  \gamma\right)  \cdot\sinh\gamma,\text{ }\left(
5J-4\right)  \sinh^{2}\gamma-\frac{1}{2}\left(  J-2\right)  \sinh
2\gamma\right]  \right]
\]
is not tighter than those of variational bounds, it can be used for studies of
low-energy physics in the LMG model e.g. for estimating the duration of the
adiabatic quantum processes in ionic traps \cite{Blatt}. There is no doubt
that it is possible to improve this lower bound. We hope to come back to these
topics in a future publication.

I am grateful to M. Fleischhauer for many fruitful and stimulating discussions.

\end{document}